\documentclass[useAMS,usenatbib]{mn2e}

\usepackage{txfonts}
\usepackage{graphicx}
\bibpunct{(}{)}{;}{a}{}{,}
\usepackage{color}
\usepackage{ulem}
\newcommand{\rrlyr} {RR~Lyr}
\newcommand{\kepl} {{\it Kepler}}
\newcommand{\puls} {$P_0$}
\newcommand{\pbla} {$P_B$}
\newcommand{\tmax} {T$_{\it max}$}
\newcommand{\kmax} {$K_{p,max}$}

\title[The historical vanishing of the Blazhko effect of RR~Lyr]
{The historical vanishing of the Blazhko effect of RR~Lyr \\ from GEOS and \kepl\,
surveys}

\author[J.F. Le Borgne et al.]{
J. F. Le Borgne,$^{1,2,3}$\thanks{E-mail: jleborgne@irap.omp.eu}
E. Poretti,$^{1,2,3,4}$
A. Klotz,$^{1,2,3}$
E. Denoux,$^{3}$
H. A. Smith,$^{5}$
\newauthor
K. Kolenberg,$^{6,7}$
R. Szab\'o,$^{8}$
S. Bryson,$^{9}$
M. Audejean,$^{10}$
C. Buil ,$^{11}$
J. Caron,$^{12}$
\newauthor
E. Conseil,$^{13}$
L. Corp,$^{3,14}$
C. Drillaud,$^{13}$
T. de France,$^{14}$
K. Graham,$^{14}$
K. Hirosawa,$^{15}$
\newauthor
A.N. Klotz,$^{3}$
F. Kugel,$^{12}$
D. Loughney,$^{16}$
K. Menzies,$^{14}$
M. Rodr\'\i guez,$^{17}$
and
P. M. Ruscitti$^{18}$\\
\\
$^{1}$Universit\'e de Toulouse; UPS-OMP; IRAP; Toulouse, France\\
$^{2}$CNRS; IRAP; 14, avenue Edouard Belin, F-31400 Toulouse, France\\
$^{3}$GEOS (Groupe Europ\'een d'Observations Stellaires), 23 Parc de Levesville, 28300 Bailleau l'Ev\^eque, France\\
$^{4}$INAF-Osservatorio Astronomico di Brera, Via E. Bianchi 46, 23807, Merate (LC), Italy\\
$^{5}$Michigan State University, Department of Physics and Astronomy, East Lansing, MI 48824, USA\\
$^{6}$Harvard-Smithsonian Center for Astrophysics, 60 Garden Street, Cambridge MA 02138, USA\\
$^{7}$Instituut voor Sterrenkunde, K.U. Leuven, Celestijnenlaan 200D, 3001 Heverlee, Belgium\\
$^{8}$Konkoly Observatory, MTA CSFK, Konkoly-Thege Mikl\'os \'ut 15-17, H-1121 Budapest, Hungary\\
$^{9}$NASA Ames Research Center, Moffett Field, Mountain View, CA 94035, USA\\
$^{10}$Observatoire de Chinon, Astronomie en Chinonais, Mairie, Place du G\'en\'eral de Gaulle, 37500, Chinon, France\\
$^{11}$Observatoire de Castanet-Tolosan, 6 place Cl\'emence Isaure, 31320, Castanet-Tolosan, France\\
$^{12}$Observatoire Chante-Perdrix, Dauban, 04150 Banon, France\\
$^{13}$AFOEV (Association Fran\c caise des Observateurs d'Etoiles Variables), Observatoire de Strasbourg 11, 
rue de l'Universit\'e, 67000 Strasbourg, France\\
$^{14}$AAVSO (American Association of Variable Star Observers), 49 Bay State Rd., Cambridge, MA 02138, USA\\
$^{15}$Variable Star Observers League in Japan (VSOLJ), 405-1003 Matsushiro, Tsukuba, Ibaraki 305-0035, Japan\\
$^{16}$The British Astronomical Association, Variable Star Section (BAA VSS), 
Burlington House, Piccadilly, London, W1J 0DU, United Kingdom\\
$^{17}$Alberdi 42 2F, 28029 Madrid, Spain\\
$^{18}$Osservatorio Astronomico B. Occhialini, Via G. Garibaldi 17, 67041 Aielli (AQ), Italy\\
}

\begin{document}

\date{Accepted ..... Received ....; in original form ....}

\pagerange{\pageref{firstpage}--\pageref{lastpage}} \pubyear{2014}

\maketitle

\label{firstpage}

\begin{abstract}
RR~Lyr is one of the most studied variable stars. Its light curve has been regularly 
monitored since the discovery of the periodic variability in 1899. Analysis of 
all observed maxima allows us to identify two primary pulsation states defined as 
pulsation over a long ($P_0$ longer than 0.56684 d) and a short ($P_0$ shorter 
than 0.56682 d) primary pulsation period. These states alternate with intervals 
of 13-16~yr, and are well defined after 1943. The 40.8\ d periodical modulations 
of the amplitude and the period (i.e. Blazhko effect) were noticed in 1916. We 
provide homogeneous determinations of the Blazhko period in the different primary 
pulsation states. The Blazhko period does not follow the variations of $P_0$ and 
suddenly diminished from 40.8\ d to around 39.0\ d in 1975. The monitoring of 
these periodicities deserved and deserves a continuous and intensive observational 
effort. For this purpose we have built dedicated, transportable and autonomous 
small instruments, Very Tiny Telescopes (VTTs), to observe the times of maximum 
brightness of RR~Lyr. As immediate results the VTTs recorded the last change of 
$P_0$ state in mid-2009 and extended the time coverage of the {\it Kepler} 
observations, thus recording a maximum O-C amplitude of the Blazhko effect at 
the end of 2008, followed by the historically smallest O-C amplitude in late 2013. 
This decrease is still ongoing and VTT instruments are ready to monitor the 
expected increase in the next few years.
\end{abstract}

\begin{keywords}
techniques: photometric – stars: individual: RR~Lyrae – stars: oscillations – 
stars: variables: RR~Lyrae.

\vspace{1cm}

\end{keywords}

\section{Introduction}
The modulation of the amplitude in luminosity and the pulsation period, known 
as Blazhko effect, is observed in numerous RR~Lyr stars \citep{blagal}. Alhough 
its theoretical explanation has not yet been determined, a considerable breakthrough 
in its interpretation has been realized because of recent space observations.
As a matter of fact, CoRoT \citep{793,eli1} and \kepl\, \citep{eli2} results
have shown that the Blazhko mechanism is not acting as a clock, thus undermining 
both the competing models based on a strict regularity: the oblique pulsator and 
the resonant nonradial pulsator. In our previous work on galactic Blazhko stars 
\citep{blagal}, we did not consider the eponym of the class, i.e., 
HD~182989$\equiv$RR~Lyr. We know that its pulsational period shows some apparent 
erratic changes \citep{pervar}. Indeed, we thought that the observation of 
\rrlyr\, through the ages deserved a detailed study and, furthermore, dedicated 
projects.

The variability of \rrlyr\, was discovered by Mrs.~Willemina P. Fleming
\citep[see][for a short but complete biography]{cannon} on the photographic plates 
of the Henry Draper Memorial in 1899 \citep{pickering}. \citet{wendell} reports 
that the first observation dates back to July 20, 1899 and the first maximum to 
September 23, 1899 \citep[HJD~2414921.675, calculated by][]{schutte}.
\citet{prager} and \citet{shapley} were the first to show the correlated 
photometric and spectroscopic variations. Fig. 5 of Shapley (1916) constitutes 
the {\it ante litteram} representation of the Blazhko effect – note that this 
definition was introduced later in the astronomical literature – because it shows 
the oscillation of the median magnitude of the ascending branch with a period of
40~d and an amplitude of 37~min.

\citet{detre43} compiled the first long list of observed maxima and reported a 
detailed investigation of the Blazhko effect \citep{balazs}. \citet{preston65} 
found that spectroscopic and photometric parameters were strongly variable in 
the Blazhko cycles of 1961 and 1963, but essentially constant in those of 1962. 
The idea of irregularities in the Blazhko period gradually grew and later extended 
photoelectric observations suggested a 4-year cycle in the Blazhko effect, with 
a minimum variability in 1971 June and a new maximum variability in 1972 April 
\citep{detreibvs}. A hundred years after the discovery of the \rrlyr\ variability, 
\citet{budapest} have reported that attempts at representing the Blazhko variations 
by longer periods have failed because they are not strictly repetitive. Extensive 
photoelectric and CCD observations obtained over a 421-day interval in 2003-2004
fixed a shorter Blazhko period of 38.8$\pm$0.1~d \citep{kk2006}. Finally,
\rrlyr$\equiv$KIC7198959 was included in the field of view of the \kepl\, space 
telescope \citep{borucki}. Thus, high-precision, continuous observations could 
be secured, first in the long cadence (29.4~min) mode and then in the short cadence 
(1.0~min) observing mode. \citet{kk2011} have reported the results in the long 
cadence mode, while \citet{molnar} and \citet{pdm} have reported the results in 
the short cadence mode. \rrlyr\, shows no particularities in metallicity, abundances 
and physical characteristics when compared with the other \kepl\, RR~Lyrae stars, 
Blazhko and non-Blazhko, observed by means of homogeneous high-resolution 
spectroscopy \citep{nemec}.

It is clear that the full comprehension of the \rrlyr\, variability is very time 
demanding since the star has to be observed over decades with time series having 
a high temporal resolution, in order to survey the pulsation period (\puls) of 
13~h, the Blazhko period (\pbla) of 40~d, and the long-term changes of both.
\begin{table}
\caption{Observers and observing instruments.}
\label{observers}
\begin{tabular}{lll}
\hline
\multicolumn{1}{c}{Observer} &
\multicolumn{1}{c}{Telescope} &
\multicolumn{1}{c}{Detector} \\
\hline
Maurice Audejean        & Reflector 320mm & CCD \\
Christian Buil         & Reflector 280mm & CCD$^{a}$ \\
Emmanuel Conseil        & Reflector 150mm & DSLR \\
Laurent Corp          & Photographic lens& CCD \\
Eric Denoux           & VTT & CCD \\
Eric Denoux           & Reflector 280mm & CCD \\
Christian Drillaud       & Refractor 70mm & DSLR \\
Thibault de France       & Refractors 60mm and 80mm & CCD \\
Thibault de France       & Reflector 130mm & CCD \\
Keith Graham          & Reflector 200mm & CCD \\
Kenji Hirosawa         & Photographic lens& DSLR \\
Alain and Adrien N. Klotz     & VTT & CCD \\
F. Kugel and J. Caron      & Photographic lens& CCD \\
F. Kugel and J. Caron      & Refractor 80mm & CCD \\
Jean-Fran\c cois Le Borgne   & VTT & CCD \\
Des Loughney          & Photographic lens & DSLR \\
Kenneth Menzies         & Reflector 317mm & CCD \\
Miguel Rodr\'\i guez      & Refractor 60mm & CCD \\
Paolo Maria Ruscitti      & Reflector 130mm & DSLR \\
Horace A. Smith and coll.     & Reflector 600mm$^{b}$ & CCD \\
\hline
\end{tabular}
$^{a}$Synthetic photometry from low resolution spectra using Shelyak
Alpy 600 spectrograph.\\
$^{b}$Telescope at the Michigan State University campus Observatory
(East Lansing, Michigan, USA) operated by H. Smith, with the help of Michigan
State University students Charles Kuhn, James Howell, Eileen Gonzales, and Aron
Kilian.
\end{table}
\section{Observations}
With the goal of making the ground-based survey of the Blazhko effect of RR~Lyr
as effective as possible, we decided to devise small, autonomous, and transportable 
photometric instruments. The basic idea was that these instruments are able to 
follow \rrlyr\ continuously, in order to obtain a reliable time of maximum 
brightness (\tmax) on as many clear nights as possible. Since the calibration of 
a photometric system would have required a major refiniment and would have increased 
the costs of the instrument (e.g. filters, standard stars observations, cooling, 
etc.), both ill-suited for the requirements of simplicity and duplicability, the 
determination of a standard, calibrated magnitude of the maximum was not pursued 
as a goal. The first observations were performed in 2008 and these are expected 
to continue for many years to come. The instruments are composed of a commercial 
equatorial mount (Sky-Watcher HEQ5 Pro Goto), an AUDINE CCD camera (512x768 kaf400 
chip) and a photographic 135-mm focal, f/2.8 lens with a field of view of 
2$^\circ$x3$^\circ$. We gave them the nickname, Very Tiny Telescopes (VTTs). Three 
such instruments have been built and used mainly in 3 different places near 
Toulouse, Castres and Caussades (R\'egion Midi-Pyr\'en\'ees, France). However they 
have been occasionally moved to several other places in France, Spain, and Italy.

The observations of the VTTs are controlled from a computer using the program
AudeLA. VTT images were obtained with no filter with an exposure time of 30~s.
Science images have been corrected with mean dark images obtained during the night 
to compensate the absence of cooling. A dark frame was obtained every five images 
on the target and the mean dark image of five individual dark frames was subtracted 
from the 25 target images concerned. From June 2008 to November 2013, 332 maxima 
were measured by VTTs with about 360\,000 images collected on 829 nights. The 
photometry of \rrlyr\, and of the comparison star HD~183383 was performed with the 
SExtractor software \citep{bertin}. The times of maximum brightness were determined 
by means of cubic spline functions with a non-zero smoothing parameter which depends 
on the number of measurements to fit and on their scatter \citep{reinsch}. The 
smoothing parameter was chosen to be large enough to avoid local maximums and so 
that the fitting curve goes through the points with zero mean residuals over a 
characteristic time interval of 5~min. The uncertainty on the time of the maximum 
was the difference between the two times corresponding to the intersection of 
the spline function with the line $y=m_{max} + \sigma/\sqrt{N-1}$, where $m_{max}$ 
is the instrumental magnitude at maximum, $\sigma$ is the standard deviation of 
the fit and $N$ is the number of measurements used in the light curve.

Alhough VTT observations are most numerous during the 2008-2013 campaign (829 
night runs compared to 938 in total), observations with other intruments are also 
included in the present study (Table~\ref{observers}). Most of these additional 
observations where carried out with classical telescopes, refractors or reflectors, 
with diameters from 55~mm to 320~mm, equipped with CCD cameras. Among these, {\it 
BVI} measurements were obtained with the 60-cm telescope at the Michigan State 
University campus observatory. However, some observations were carried out with 
digital photographic cameras (DSLR). As a new approach to spectrophotometry, 
synthetic photometry was performed on low-resolution spectra using a Shelyak Alpy 
600 spectrograph mounted on a 280-mm diameter reflector. Low-resolution spectra 
($R$=600) were obtained through a wide slit and calibrated in flux by means of 
spectrophotometric standard stars. Photometry is then performed by integrating 
the spectra in Johnson filter bandpasses.

\section{GEOS data base}

The Groupe Europ\'een d'Observations Stellaires (GEOS) RR~Lyr data base
\footnote{http://rr-lyr.irap.omp.eu/dbrr/dbrr-V1.0\_08.php?RR\%20Lyr} (Le Borgne 
et al. 2007) is a collection of published maxima of galactic RR~Lyr stars which 
contains 2245 maxima of RR~Lyr itself (up to 2013 December 5). In Appendix A, we 
give the list of references used to build the GEOS data base for RR~Lyr. In our 
analysis, we did not consider uncertain maxima and those noted by the authors
as normal, created from observations drawn from many different individual maxima. 
This is because they could be referred to a wrong epoch if an inaccurate \puls\, 
value were to be used. The normal maximum could also be the arbitrary epoch of 
an ephemeris. This is the case of the first ephemerides of \rrlyr\, \citep{prager,
shapley,sanford} that were calculated using JD~2414856, i.e., the date of the 
first observation, rather than JD 2414921, i.e., the date of the first maximum.
Moreover, a normal maximum masks the Blazhko effect because it averages observations 
on a large interval of time.

We added 692 new photographic and photoelectric maxima observed by L.~Detre in 
the years 1944-1981. The list of $\sim$7000 measurements reported by \citet{Szeidl97} 
was scanned and digitalized by means of a semi-automatic procedure. These 
measurements are now available on electronic form on the Konkoly Observatory web 
site. We also evaluated the differences between 278~\tmax\, observed simultaneously 
in $B$ and $V$ filters. No systematic effect was detected. 67 per cent of the 
\tmax\ differences are within the interval from $-0.0016$ to $+0.0016$~d, 
symmetrically distributed with respect to 0.000~d. The frequency analysis did not 
detect any periodicity in the \tmax\, differences. Finally, we measured 25 \tmax\, 
values from the original observations collected at Michigan University in the 
framework of the 2003-04 campaign \citep{kk2006}.

\section{\puls\, variations of RR~Lyr over more than one century}

We analysed 3975 \tmax\, (obvious outliers were removed) spanning 114~yr and we 
calculated the linear ephemeris from a least-squares fitting of all them:
\begin{equation}
{\rm HJD\,\,\, Max}\ = 2414921.7746 + 0.566835616 \cdot {\rm E}
\label{cst}
\end{equation}

\begin{figure*}
\resizebox{\hsize}{!}{\includegraphics{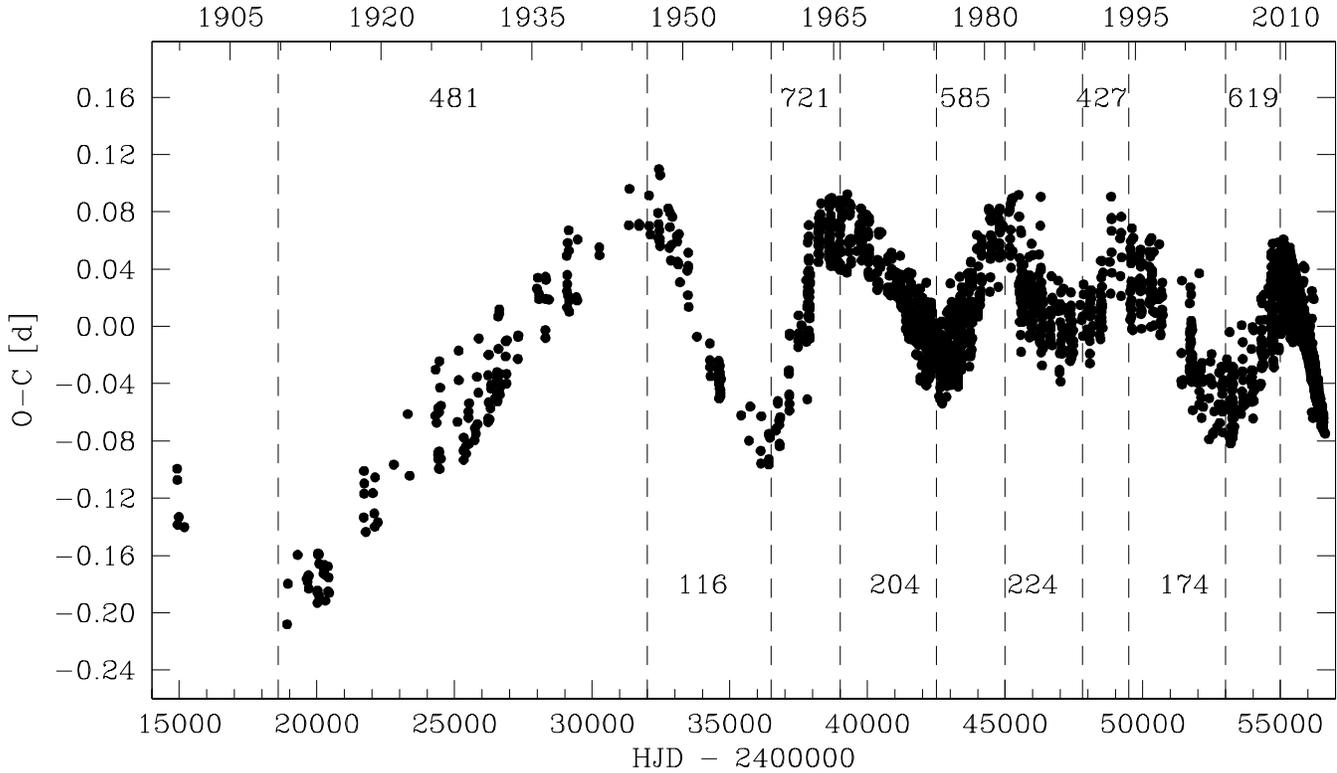}}
\caption{Historical behaviour of the period variations of RR~Lyr.
The numbers are the last three digits of the pulsational period
calculated in each interval.}
\label{cstperiod}
\end{figure*}

In Appendix B, we give the list of \tmax used to calculate equation~\ref{cst} 
and used in the subsequent analysis. The purpose of using the above ephemeris was 
to detect large changes in the \puls\, value. Indeed, the plot of the O-C 
(observed minus calculated \tmax) values clearly pointed them out 
(Fig.~\ref{cstperiod}). Therefore, we subdivided the 114~yr of observations of 
\rrlyr\, into several time intervals, following the \puls\, changes regardless 
of the observing technique. Table~\ref{history} lists the actual \puls\, values 
in each interval.

The Blazhko period contributes greatly to an increase in the O-C scatter beyond
that attributable to the observing technique alone. The varying thickness in 
Fig.~\ref{cstperiod} also suggests a variable amplitude. To study the behaviour 
of \puls, we calculated a linear ephemeris in each interval of Table~\ref{history} 
this time by dividing the \tmax on the basis of the observing technique (visual, 
photographic, photoelectric, or CCD) or instrument (VTTs, \kepl). These subsets 
supplied independent values of \puls\, (Table~\ref{history2}) in good agreement 
with those of the whole time interval (Table~\ref{history}). The uncertainties 
on \puls\, (as well as those on the Blazhko period \pbla, see below) are the formal 
error bars derived from the least-squares fittings. About the first long time 
interval, we preferred to use the visual maxima instead of the photographic maxima 
because the visual technique implied the survey of the star for several hours, 
while the photographic technique often recorded a few measurements only. The visual 
maxima are very useful to fill the long gap between 1982 and 2000, not covered 
by photoelectric observations before many amateur astronomers upgraded their 
instrumentation to CCD detectors. In addition to the time intervals listed in 
Table~\ref{history2}, we note that visual maxima provided \puls\, values in 
excellent agreement with the photoelectric and CCD ones in the intervals JD 
2439000-2442500, 2442500-2445000, and 2455000-2456200.

\begin{table}
\caption{Pulsation periods \puls\, in the intervals of Fig.~\ref{cstperiod}.
Error bars on the last digits are between brackets.}
\label{history}
\centering
\begin{tabular}{ll r l}
\hline\hline
\noalign{\smallskip}
\multicolumn{1}{c}{Years} &
\multicolumn{1}{c}{Julian Days} &
\multicolumn{1}{c}{N$_{\rm max}$} &
\multicolumn{1}{c}{Pulsational period} \\
&\multicolumn{1}{c}{[JD-2400000]} & &
\multicolumn{1}{c}{\puls\,[d]} \\
\noalign{\smallskip}
\hline
\noalign{\smallskip}
1899-1908 & 14921-18000 &  5 & 0.5668 (1) \\
1908-1946 & 18000-32000 & 138 & 0.5668 481(3) \\
1946-1958 & 32000-36500 & 70 & 0.5668 116(9) \\
1959-1965 & 36500-39000 & 164 & 0.5668 721(17) \\
1965-1975 & 39000-42500 & 418 & 0.5668 204(4) \\
1975-1981 & 42500-45000 & 324 & 0.5668 585(10) \\
1982-1989 & 45000-47800 & 205 & 0.5668 224(12) \\
1989-1995 & 47800-50000 & 103 & 0.5668 427(17) \\
1995-2003 & 50000-53000 & 132 & 0.5668 174(10) \\
2004-2009 & 53000-55000 & 188 & 0.5668 619(14) \\
2009-2013 & 55000-57000 &2228 & 0.5667 975(4) \\
\noalign{\smallskip}
\hline
\end{tabular}
\end{table}

\begin{table*}
\caption{Pulsation \puls\, and Blazhko \pbla\, periods calculated from homogenous subsets.
Error bars on the last digits are between brackets.
}
\label{history2}
\centering
\begin{tabular}{l l r ll c}
\hline\hline
\noalign{\smallskip}
\multicolumn{1}{c}{Julian Days} &
\multicolumn{1}{c}{Method} &
\multicolumn{1}{c}{{\bf Number}} &
\multicolumn{1}{c}{Pulsation period} &
\multicolumn{1}{c}{Blazhko} &
\multicolumn{1}{c}{Blazhko O-C} \\
\multicolumn{1}{c}{[JD-2400000]} &
& 
\multicolumn{1}{c}{{\bf of \tmax}} &
\multicolumn{1}{c}{[d]} &
\multicolumn{1}{c}{period [d]} &
\multicolumn{1}{c}{amplitude [d]} \\
\noalign{\smallskip}
\hline
\noalign{\smallskip}
19635-27313 & visual    &  75 & 0.5668 470(5) & 40.89(3) & 0.018(3) \\
32062-33455 & photog.   &  27 & 0.5668 193(45)& 40.86(28) & 0.014(3)\\
33505-36457 & photoel.   &  39 & 0.5668 200(9) & 40.94(28) & 0.005(2) \\
36674-38996 & photoel.   & 160 & 0.5668 734(17)& 40.88(10) & 0.018(2)\\
39008-42405 & photoel.   & 342 & 0.5668 205(4) & 41.15(3) & 0.014(1)\\
42504-44822 & photoel.   & 162 & 0.5668 583(13)& 38.92(12) &0.013(2)\\
45493-47779 & photoel.   &  14 & 0.5668 271(26) & \multicolumn{2}{c}{Too few points} \\
45131-47982 & visual   & 193 & 0.5668 239(13) & 39.02(20) & 0.007(3) \\
48012-50000 & visual   &  98 & 0.5668 427(18) & 39.03(10) & 0.018(3)\\
50224-52926 & visual    & 126 & 0.5668 1703(10) & 39.06(11) & 0.012(2) \\
52915-54733 & CCD &      40 & 0.5668 621(28) & 39.00(5) & 0.024(2) \\
54652-55000 & VTTs     &  69 & 0.5668 589(14) & 39.39(5) & 0.026(2) \\
55276-56390 & {\it Kepler} & 1815 & 0.5667 953(4) & 38.84(2) & down to 0.009 \\
55000-56624 & VTTs    & 264 & 0.5668 024(11) & 38.91(7) & down to 0.006\\
\noalign{\smallskip}
\hline
\end{tabular}
\end{table*}

After an initial change, which the few observed maxima place around 1910, the 
period of \rrlyr\, was constant for about 36 yr. Then, around 1946, it started 
a series of suddden changes, on a timescale of few years. The O-C pattern shows 
jumps from long values (O-C's from negative to positive values, with maxima values 
reached on 1946.5, 1965.7, 1982.1, 1994.4, and 2009.5) to short values (O-C's 
from positive to negative values, with minima values reached on 1958.8, 1975.2, 
1989.8, and 2004.0). Table~\ref{history} lists the computed values of \puls\, 
after any observed change and the last three digits of \puls\, are also noted in 
Fig.~\ref{cstperiod}. The minimum difference between a long and a short value is 
between JD~2445000 and 2450000. One of the changes of state is well covered by VTT
observations: when considering the \tmax before JD~2455000 the period is a long 
one (0.56686~d), after a short one (0.56680~d, see Table~\ref{history2}).

We also note that there is a sort of semiregular cadence separating two consecutive 
O-C maxima or two consecutive O-C minima, about 13--16 years and that the \puls\, 
is still decreasing: if we consider the maxima after the change at JD~2455000 
only, the period is below 0.5668000~d (Tables~\ref{history} and~\ref{history2}).

\section{\pbla\, variations of RR~Lyr over more than one century}

Because the pulsation period of \rrlyr\, was undergoing changes from two different
states, the question of whether these changes effect the Blazhko effect immediately 
arose. This point could be very important in understanding the relation between 
the two periods. It required the analyses of all the previous observations by a 
homogenous procedure, because often the \pbla\, values were determined in different 
ways or also simply assumed from previous works.

\begin{figure}
\resizebox{\hsize}{!}{\includegraphics{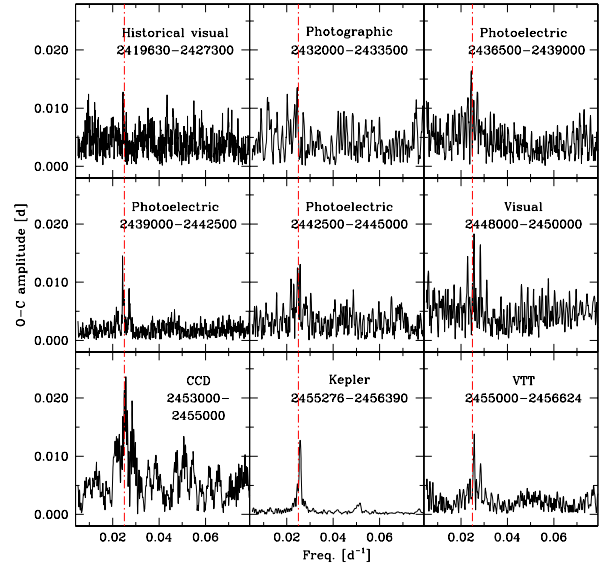}}
\caption{The spectra show the amplitudes and the frequencies of the Blazhko effect
in the O-C values of \rrlyr\, as measured with different techniques through the
ages. The (red) vertical line is at $f$=0.025~d$^{-1}$ ($P$=40.0~d).}
\label{spectra}
\end{figure}

Therefore, we performed the frequency analysis of the O-C values obtained from 
the linear fits used to determine the \puls\, in each subset (Table~\ref{history2}). 
We used the iterative sine-wave fitting method \citep{vani} and we present the 
results as amplitude spectra for the sake of clarity. Figure~\ref{spectra} shows
the application of the method to several subsets covering a time interval of one 
century. We can clearly see that the amplitude is variable and that \pbla\, jumps 
from left to right of the mark at 0.025~d$^{-1}$ (i.e., 40.0~d) around 1975. The 
\pbla\, values obtained from the highest peaks were refined by means of the MTRAP
code \citep{mtrap} and the final values are listed together with error bars and 
amplitudes in the last two columns of Table~\ref{history2}. It is quite evident 
that \puls\, and \pbla\, changed in a way completely uncorrelated with each other
(see Sect.~\ref{disc}).

\section{The \kepl\, data}

\begin{figure}
\resizebox{\hsize}{!}{\includegraphics{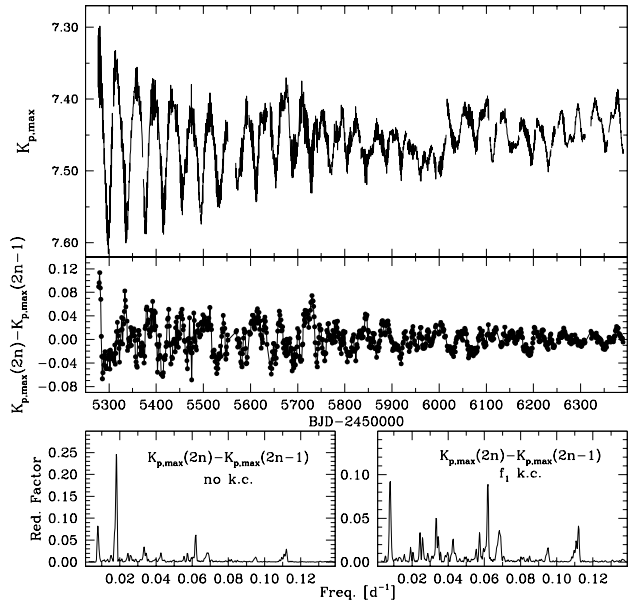}}
\caption{The period doubling effect in the magnitudes at maximum brightness of 
\rrlyr\, observed with \kepl. Top panel: \kmax\, values. Consecutive values 
are connected. Middle panel: plot of the differences between two consecutive 
cycles in the sense odd-epoch (2n) minus even-epoch (2n-1). Bottom 
panels: power spectra of the values of the middle panel, with no known constituent
(left) and with $f_1$=0.018~d$^{-1}$ (right) as known constituent.
}
\label{doubling}
\end{figure}

\kepl\, data allowed us to combine the analysis of \tmax\, variations with another 
specific Blazhko characteristic, i.e., the changes of the magnitude at the maximum 
brightness (\kmax). We determined \tmax\, and \kmax\, from the original \kepl\, 
data by means of the same procedure used for VTT data. We used the Q5-Q16 
short-cadence data acquired from 2010 March 20 to 2013 April 3. The analysis of 
the almost continuous succession of observed maxima already pointed out a totally 
new and prominent feature, i.e., the alternation of higher and lower maxima 
\citep[i.e. period doubling; see Fig.~4 in ][ for a clear example]{szabo2010}.
A theoretical background has been proposed for this new phenomenon \citep{molnar2011}.
We investigated the regularity of this effect all along the time interval of the 
\kepl\, observations (Fig.~\ref{doubling}). The scatter due to the period doubling 
effect is always noticeable (top panel). The amplitudes are variable and the 
largest amplitudes are not related to a particular phase of the Blazhko effect, 
because the related large scatter is observed at both the maximum and at the 
minimum values of \kmax. Moreover, there are Blazhko cycles where the period 
doubling effect is always very noticeable, as that from BJD 2455710 to 2455750. 
There is also a damping of the effect toward the end of observing time, when 
also \kmax\, variations have a small amplitude.

As a new contribution to the characterization of the period doubling effect, we 
calculated the differences between the \kmax\, value of an even (2$n$) epoch and 
that of an odd (2$n-1$) epoch. These differences are both positive and negative 
(middle panel) and this implies that the highest \kmax\, changes from an odd 
epoch to an even one. In this plot the highest maxima or the deepest minima are 
separated by a time interval corresponding to the characteristic period of the 
switching from an odd epoch of high \kmax\, to an even one. Moreover, more rapid 
fluctuations are also visible. It is worth analysing these time series to search 
for periodicities in the switching process. The iterative sine-wave fitting method 
\citep{vani} is well suited to disentangle such periodicities because it allows 
the detection of the components of the light curve one by one. Only the values 
of the detected frequencies (known constituents) are introduced in each new search, 
while their amplitudes and phases are recalculated for each new trial frequency. 
In such a way, the exact amount of signal for any detected frequency is always 
subtracted. In the first power spectrum of the \kmax\, differences (bottom-left 
panel), the highest peak was at a low frequency, $f_1$=0.018~d$^{-1}$, 
corresponding to $P$=55.6~d, i.e., about 98~\puls. After introducing it as a 
known constituent, we could identify a higher frequency, $f_2$=0.062~d$^{-1}$ 
(bottom-right panel), corresponding to $P$=16.1~d, i.e., about 28.5~\puls. The 
peak close to $f$=0.0~d$^{-1}$ appearing in both spectra is a result of the very 
long-term effect (see below).

\begin{figure}
\resizebox{\hsize}{!}{\includegraphics{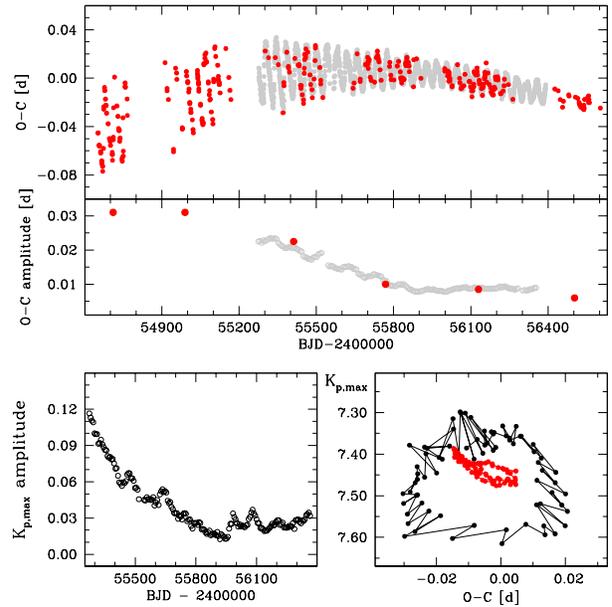}}
\caption{The changes in the Blazhko effect of \rrlyr\, from 2008 to 2013.
Top panel: VTT (red filled circles) and \kepl\, (grey circles) O-Cs showing the
strong decrease in amplitude.
Middle panel: O-C half amplitude, same symbols. VTT values are taken from the 
yearly mean curves (Fig.~\ref{ocampli}).
Bottom panel (left): \kmax\, (half) amplitude, \kepl\, data.
Bottom panel (right): the first (big) and last (small) Blazkho cycles observed 
with \kepl.
}
\label{patata}
\end{figure}

The \kepl\, data make it possible for us to study in detail the cycle-to-cycle
variations of the Blazhko effect in \rrlyr. To do this, we calculated a 
least-squares fit of the observed O-C and \kmax\, values on sliding boxes of 56~d 
shifted from each other by 8~d. This procedure returns the values of the amplitudes 
of the O-C and \kmax\, variations for each box (Fig.~\ref{patata}). The general 
trend for both amplitudes is a slow decrease. In particular, the O-C curve does 
not seem to have reached the final minimum at the end of \kepl\, observations
(middle panel), while the \kmax\, amplitude curve seems to start to increase again 
after a shallow minimum at JD 2459000 (bottom-left panel). The extreme changes 
in the Blazhko effects are sketched by the shrinking of the close curve connecting 
O-C and \kmax\, values (bottom-right panel). The alternation of low and high 
maxima also twists the regular shape of the close curve, adding a new model to 
the already variegated collection \citep{blagal}. Modulations are clearly visible 
both in the O-C and in the \kmax\, amplitudes (Fig.~\ref{patata}). Combined with 
the long-term trend, they produce the cycle-to-cycle variations of the Blazhko 
effect. We performed the frequency analysis of the time series of the O-C and 
\kmax\, amplitudes to search for periodicities by means of a sinusoidal fit and 
a parabolic trend. The two power spectra are characterized by broad structures 
with different highest peaks at low frequencies. The inconsistency between the 
two results does not support a reliable identification of real periodicities in 
the changing shape of the Blazhko effect.

\section{VTT data}

\begin{figure}
\resizebox{\hsize}{!}{\includegraphics{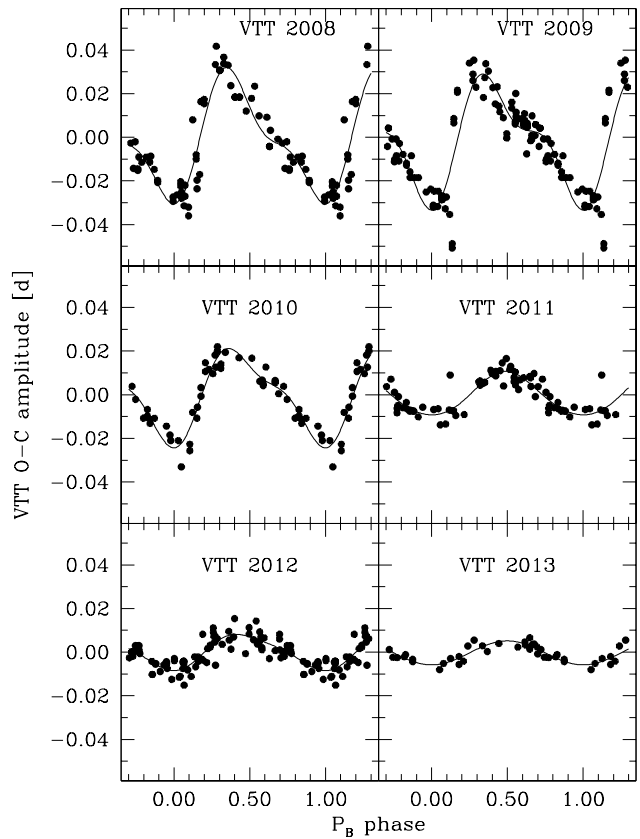}}
\caption{The decreasing amplitude of the O-C curves in the VTT data from 2008 to 2013.
}
\label{ocampli}
\end{figure}
The first VTT \tmax\, was observed on 2008 July 4 simultaneously with two 
instruments. Since then the regular survey of \rrlyr\, has yielded 55, 78, 42, 
56, 75, and 27 \tmax\, from 2008 to 2013, respectively. To analyse VTT data we 
performed a least-squares fit of the O-C values determined each year. Indeed, 
ground-based observations of RR~Lyr are concentrated in a few months of each year, 
covering about two consecutive Blazhko cycles. The variations between the Blazhko 
cycles in a given year are very small and consequently the folded curves of O-C 
over the Blazhko cycle are quite representative of the behavior in each year. 
Indeed, the continuous decline in amplitude and the changing shape of the O-C 
curve is well reproduced and variations have already become noticeable from one 
year to the next already (Fig.~\ref{ocampli}). In particular, note how the curve 
becomes more sinusoidal with the decreasing O-C amplitude.

By adding the VTT points to \kepl\, ones (Fig.~\ref{patata}, middle panel), we
can state that the maximum O-C amplitude has been reached well before the space
observations began. The CCD determinations preceeding the VTT observations 
(40~\tmax\, between JD 2452915 and 2454733) supply an O-C amplitude of 0.024~d 
(Fig.~\ref{spectra}), while the visual \tmax~s collected before CCD ones supply 
smaller amplitudes, i.e., 0.018$\pm$0.003~d and 0.012$\pm$0.002~d 
(Table~\ref{history2}). From these values we can infer that a minimum O-C 
amplitude occurred when only visual \tmax~s were available, in 1984-85 (tentatively
around JD~2446000) followed by an increase that reached its maximum (0.031~d) 
at the beginning of the VTT observations, near the end of 2008 (JD~2454800, see 
Fig.~\ref{ocampli}, top panels). The subsequent decrease is still ongoing 
(bottom panels). Indeed, the O-C amplitude recorded by the VTTs until the end 
of 2013 (0.006~d) is even smaller than that of the last \kepl\, data 
(Fig.~\ref{patata}, middle panel).

\section{Discussion}\label{disc}

\begin{figure}
\resizebox{\hsize}{!}{\includegraphics{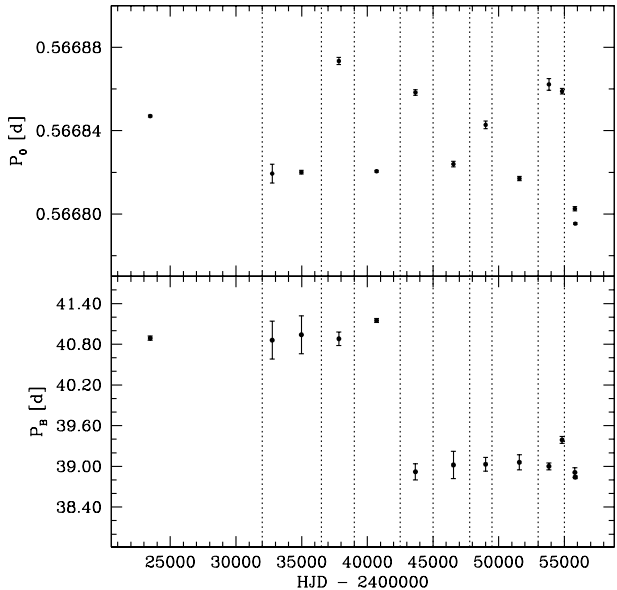}}
\caption{Uncorrelated variations of the pulsation period \puls\, (top panel) and
of the Blazhko period \pbla\, (bottom panel). Time interval are the same 
as in Fig.~\ref{cstperiod}. 
}
\label{correl}
\end{figure}

The VTT observations started 624~d before the first \kepl\, observation in 
short-cadence mode, and they are still continuing, while the last \kepl\, \tmax\,
was obtained in 2013, in early April. The re-analysis of the \tmax\, epochs 
listed in the GEOS data base allowed us to reconstruct the changes in the 
pulsational period of \rrlyr. We could establish the existence of two states
characterized by the pulsation over a long \puls\, (longer than 0.56684~d) and 
over a short \puls\, (shorter than 0.56682~d). The history begins with a long 
\puls\, status that lasted from 1910 to 1943. After this, the two states alternate 
more frequently and usually long states last much less than the short states. 
Since 1943 the same state seems to re-appear after a time interval of 13-16~yr 
(Fig.~\ref{cstperiod}). The frequency analysis of the O-C values since 1950 shows 
the highest peak at 14~yr. The last cycle began in 2003 with a long \puls\, and 
the VTTs recorded that it switched into a short (the shortest, actually) one in 
2009, and that is still running. The cyclic shift of \puls\, amounts on average 
to $\Delta P_0=4\cdot10^{-5}$~d (Table~\ref{history}) and hence 
$\Delta P_0/P_0=7\cdot10^{-5}$.

We also determined the periods of the Blazhko effect by means of an homogenous 
technique. We can provide a new reliable chronological set of values since the 
beginning of the twentieth century, replacing the values reported by each author 
on the basis of different methods of analysis or simply adopting values reported 
in the literature \citep[see Table~6 in][for a detailed list]{kk2006}. The 
alternate states of \puls\, do not have a counterpart in the variations of \pbla\, 
(Fig.~\ref{correl}). Actually, when comparing the whole set of the new 
determinations of \pbla\, we can argue that \pbla\, suddenly changed in 1975 
(around JD~2442500, Fig.~\ref{correl}), much earlier than that reported by 
\citet{kk2006}. It was around 41~d until 1975 and since then shortened to 39~d; 
the shift from one side to the other of the 40~d mark is also visible in 
Fig.~\ref{spectra}. The corresponding rate $\Delta P_B/P_B=0.05$ is three orders 
of magnitude larger than that observed for \puls. Correlated and anticorrelated 
changes of \puls\, and \pbla\, were observed in Blazhko RRab stars: RW~Dra 
\citep{firma}, XZ~Dra \citep{xzdra}, XZ~Cyg \citep{xzcyg}, RR~Gem \citep{rrgem}, 
RV~UMa \citep{rvuma}, M5 stars \citep{m5}, RZ~Lyr \citep{rzlyr}, and Z~CVn 
\citep{blagal}. \rrlyr\, shows alternate states of long and short \puls\, combined 
with the decreasing \pbla. Taking into account that changes of \puls\, and \pbla\, 
also occurred at different epochs, it seems that \rrlyr\, adds another kind of 
relation between the two periods describing the light curves of Blazhko stars.

The combination of \kepl\, and VTT data supplies us with a clear picture of the 
vanishing of the Blazhko effect. The space telescope continuously monitored the 
monotonic long-term decrease, proving that small-scale modulations, lasting from 
2 to 4 \pbla, are also visible in the O-C values. The VTTs have allowed us to 
assess that the decline in amplitude started in 2008 and it is still ongoing. 
The plot of the O-C amplitude (Fig.~\ref{patata}, middle panel) covers about 
5~yr and it shows the continuous decrease. Hence, it does not support the action 
of a 4-yr modulation cycle of the Blazhko effect \citep{detreibvs}. We also note 
that the minimum full-amplitudes of the \tmax\, and \kmax\, variations observed 
with VTTs and \kepl\, (0.012~d and 0.04~mag, respectively) are about half those 
recorded in the 1971 minimum (0.020~d and 0.07~mag). Therefore, it seems evident 
that we are observing the historical minimum level of the Blazhko effect. Such 
a small value was observed perhaps only during the sharp decrease after the O-C 
maximum in 1950 (Fig.~\ref{cstperiod}), but the event is poorly covered because 
of the small numbers of \tmax s. Another minimum O-C amplitude was perhaps
observed around 1985, but on this occasion we only obtain very scattered visual 
\tmax. We note that these three minimum O-C amplitudes occurred when \puls\,
was in the short state.

Combined with the Blazhko effect, the period doubling is making \rrlyr\, still 
more intriguing. The analysis of the \kepl\, short-cadence data was helpful in 
understanding this new effect. We can verify that this effect does not seem to 
be related to any particular Blazhko phase and can be observed at any time in 
the data. We find two clear periodicities describing a long-time (55.6~d) and a 
short-time (16.1~d) switch between the epochs (odd or even) of the higher \kmax. 
Both these periodicities are not obviously related with \pbla. We can just note 
that 55.6/38.8=1.43, roughly similar to the occurrence of half-integer values,
but the 1.5 value is not matched. Moreover, here we are dealing with periods, 
while the period doubling effect can be represented by means of half-integer 
values of the pulsational frequency ($f/2, 3/2f, 5/2f$,~...).

\section{Conclusions}

The most promising mechanism that can explain the Blazhko effect is the
$9P_9$=2\puls\, resonance between the ninth overtone and the fundamental mode,
also capable of producing the period doubling effect \citep{buchler}. A recent 
new explanation is based on the transient excitation of the first overtone radial 
mode \citep{gillet}. The signature of this mode has already been found in the
Q5-Q6 \kepl\, data \citep{molnar} and the analysis of other datasets is ongoing. 
The results described here supply a complete overview of the behaviour of the 
Blazhko effect and of the pulsation content of \rrlyr\, since its discovery 114~yr 
ago, thus putting time constraints on the Blazhko mechanism. In particular, the 
completely different behaviours of the \puls\, and \pbla\, changes suggest that 
they are not coupled in a direct way.

The VTT monitoring complemented the \kepl\, one and allowed us to follow the 
historical minimum amplitude of the Blazhko effect. The previously suggested 4-yr 
cycle does not seem effective in fitting the cycle-to-cycle and the long-term 
variations. The alternation of the long and short \puls\, with a semiregular 
timescale of 14~yr stresses the necessity to collect long series of light maxima 
in a continuous way. Because of the new \kepl\, orientation, VTTs are probably 
the only instruments that can monitor the expected regrowth of the Blazhko effect 
and to measure the new \puls\, in the coming years.

These insights lead new contributions to the description of the pulsation of \rrlyr.
They corroborate our decisions to maintain an updated data base of \tmax\, of 
\rrlyr\, stars \citep{pervar}, to monitor Blazhko stars with modern instruments
\citep{blagal}, and to have started the project to continuously monitor \rrlyr\, 
itself. These facts support us in our aim to continue and to improve the VTT 
project for several more years.

\section*{Acknowledgments}
Funding for the \kepl\, Discovery Mission is provided by NASA's Science Mission
Directorate. The \kepl\, Team and the \kepl\, Guest Observer Office are recognized 
for helping to make this mission and these data possible. EP acknowledges 
Observatoire Midi-Pyr\'en\'ees for the two-month grant allocated in 2013 October 
and November which allowed him to work at the Institut de Recherche en Astrophysique 
et Plan\'etologie in Toulouse, France. The present study has used the SIMBAD data 
base operated at the Centre de Donn\'ees Astronomiques (Strasbourg, France) and 
the GEOS RR~Lyr data base hosted by IRAP (OMP-UPS, Toulouse, France). This study 
has been supported by the Lend\"ulet-2009 Young Researchers Program of the 
Hungarian Academy of Science, the Hungarian OTKA grant K83790 and the KTIA 
URKUT\_10-1-2011-0019 grant. The research leading to these results has received 
funding from the European Community's Seventh Framework Programme (FP7/2007-2013) 
under grant agreement no. 269194 (IRSES/ASK). RSz was supported by the J\'anos 
Bolyai Research Scholarship of the Hungarian Academy of Sciences. The basic ideas 
of the VTT project have been sketched during several GEOS meetings and fruitful 
discussions with R.~Boninsegna, M.~Dumont, J.~Fabregat, F.~Fumagalli,
D.~Husar, J. Remis, J.~Vandenbroere, and J.M.~Vilalta are gratefully acknowledged.

\appendix
\section{References to published times of maximum}
The historical study of the variations of the pulsating period and of the Blazhko 
effect period of RR~Lyr uses determinations of times of individual maxima reported 
in numerous publications since the early twentieth century. We list here the 
references of the publications available, to the best of our knowledge. As said 
in the text, the reference \citet{Szeidl97} does not contain determinations of 
times of maximum, but the measurements from which we determined the individual 
times of maxima.

\section{Historical maximum list}

Table \ref{maxima} lists the times of maximum of RR~Lyr used to determine the 
historical linear ephemeris.\\
The columns give the following: \\
 HJD = Heliocenric Julian Day \\
 Uncert. = Estimated uncertainty on the time of maximum \\
 O-C : Observed time of maximum minus calculated time of maximum \\
 E : Cycle number used in the linear ephemeris to obtain the calculated time of maximum \\
 Reference : Paper where the tabulated time of maximum is reported \\
 Observer: Name of the observer, if specified in the Reference \\
 Method of observation: visual (vis), photographic (pg), pe (photolectric), ccd (CCD), 
            dslr (digital photographic camera) \\
 Comment : supplementary information. \\
The complete table is available online as supporting information.

\begin{table}
\caption{lists the times of maximum of RR~Lyr used to determine the historical linear ephemeris.}
\label{maxima}
\begin{tabular}{llllllll}
\hline
\multicolumn{1}{c}{HJD} &
\multicolumn{1}{c}{Uncert.} &
\multicolumn{1}{c}{O-C} &
\multicolumn{1}{c}{E} &
\multicolumn{1}{c}{Reference} &
\multicolumn{1}{c}{Observer} &
\multicolumn{1}{c}{Method} &
\multicolumn{1}{c}{Comment} \\
\hline
2414921.6750 & 0.01 & -0.0996 &   0 & Wendell,1909, Wendell,1914 &  O.C. Wendell   &   vis & \\
2414925.6350 & 0.01 & -0.1074 &   7 & Wendell,1909, Wendell,1914 &  O.C. Wendell   &   vis & \\
2414938.6410 &    & -0.1387 &  30 & Wendell,1909, Wendell,1914 &  O.C. Wendell   &   vis & \\
2414984.5600 &    & -0.1334 &  111 & Wendell,1909, Wendell,1914 &  O.C. Wendell   &   vis & \\
2415184.6460 &    & -0.1403 &  464 & Wendell,1909, Wendell,1914 &  O.C. Wendell   &   vis & \\
2418919.4580 & 0.01 & -0.2082 & 7053 & Hertzsprung,1922      &  E. Hertzsprung  &   pg & \\
2418944.4270 & 0.01 & -0.1800 & 7097 & Hertzsprung,1922      &  E. Hertzsprung  &   pg & \\
\hline
\end{tabular}
\end{table}

\label{lastpage}

\end{document}